\documentstyle[twoside,epsf,12pt,a4]{article}
\newcommand{\insertplot}[1]{
       \begin{center}
       \leavevmode
       \epsfysize=9.0cm\epsfbox{#1}
       \end{center} }
\catcode`\@=11 \relax
\def\ps@burst{%
        \def\@oddhead{}  \def\@evenhead{}
        \def\@oddfoot{%
{\sl GSI-Preprint 95-47\hfil}
        } %END defining \@oddfoot
        \let\@evenfoot\@oddfoot
} %END defining \ps@firstpage
\catcode`\@=12 \relax
\newcommand {\NP}[1]{ Nucl. Phys. {\bf #1}}

\newcommand {\PR}[1]{ Phys. Rev. {\bf #1}}

\newcommand {\PRE}[1]{ { Phys. Reports} {\bf #1}}
\newcommand {\PL}[1]{ { Phys. Lett.} {\bf #1}}

\newcommand{\tauj}{\tau^{\ }_{\rm j}}
\newcommand{\taui}{\tau^{\ }_{\rm i}}

\begin{document}
\thispagestyle{burst}
\begin{center}
{\Large \bf Multifragmentation calculated with \\
relativistic forces}
\end{center}

\vspace*{9mm}

\hspace*{9mm}\begin{minipage}[t]{12cm}
{\large
H. Feldmeier$^1$, J. N\'emeth$^2$ and G. Papp$^{1,2\dagger}$
}\\[5mm]
$^1$Gesellschaft f\"ur Schwerionenforschung, D-64220, Darmstadt, \\
\phantom{$^1$}Germany\\
$^2$Institute for Theoretical Physics of E\"otv\"os University,\\
\phantom{$^2$}H-1088 Budapest,  Hungary

\vspace*{6mm}{\it

Submitted to {\it Heavy Ion Physics}
}\\[9mm]
    \hrule height 0.4pt
    \vskip 8pt

{\bf Abstract.}
A saturating hamiltonian is presented in a relativistically
covariant formalism. The interaction is described by scalar and vector mesons,
with coupling strengths adjusted to the nuclear matter. No explicit density
dependence is
assumed. The hamiltonian is applied in a QMD
calculation to determine the fragment distribution in O + Br collision
at different energies (50 -- 200 MeV/u) to test the applicability of
the model at low energies. The results are compared with experiment and
with previous non-relativistic calculations.
\\[9mm]
    \vskip 10pt
    \hrule height 0.4pt
\end{minipage}

\newpage

\section*{\large\bf 1. Introduction}

In molecular dynamics simulations of
heavy-ion collisions at intermediate energies (300 MeV/u -- 2 GeV/u)
it is still an open question
question, which kind of forces should be used for the long range part of the
nucleon-nucleon interaction which builds up the mean field. The Skyrme forces,
which work very well for lower energies are not relativistically invariant.
There are attempts to extend non-relativistic two-body potentials
to higher velocities by requiring approximate Lorentz
invariance~\cite{Fel1,Schm89,Stac76,Bodm80}.
Our aim is to derive from a
relativistic lagrangian of the Walecka type a covariant Lagrange function
for point-like particles
which is suited for molecular dynamics and still has the
saturation property of the original lagrangian. It should then be
applicable in transport theories as for example
QMD or BUU and should give reasonable results even at lower energies (50 --
200 MeV/u).

In Section~2 the relativistic Hamilton function is derived in the small
acceleration approximation \cite{Weber} from a many-body Lagrange function
and it is expressed in terms of positions
and canonical momenta of the particles.
In Section~3 the QMD procedure is described, while in
Section~4 we present our QMD calculations for fragment distributions in
O + Br collisions at different energies and compare the calculated results with
the experimental data. A few years ago the analogue calculation was
done~\cite{kn:judqmd} with
the Skyrme interaction, which is explicitly density dependent. It is
interesting to compare
those results with the new forces presented in this paper in order to see how
far they can describe saturation and low energy phenomena.

The intention is, however, to go to intermediate energies, where QMD is more
appropriate because quantum aspects like the Pauli principle or the uncertainty
relation are expected to be less important.

\section*{\large\bf 2. Relativistic equations of motion for nucleons}

In this section we are deriving the relativistic equations of motion
for the mean 4-positions and the mean 4-momenta of the nucleons.
We start from a field theoretical lagrangian of the following type
%---------------------------------------------------------------------------
\begin{eqnarray}\label{Lagr1}
{\cal L}(x) &  = & \bar{\psi}(x) \Big( \gamma^\alpha i\partial_\alpha
		- m^*\left( \phi(x)\right) \Big)
\psi(x) \nonumber \\
& - & g_{\rm v} \bar{\psi}(x) \gamma^\alpha \psi (x)  A_\alpha (x) \nonumber \\
& - & \frac{1}{2} \phi(x) (\partial_\alpha \partial^\alpha + \mu^2_{\rm s})
\phi (x) \nonumber \\
& + & \frac{1}{2} A^\alpha(x) (\partial_\beta \partial^\beta + \mu^2_{\rm v})
A_\alpha (x) \ ,
\end{eqnarray}
%---------------------------------------------------------------------------
where we allow for two different ways to couple the scalar
field~\cite{Lindner}. If
$m^*(\phi) = m - g_{\rm s} \phi$ we obtain the Walecka lagrangian, whereas for
$m^*(\phi) = \left( m + g_{\rm s} \phi \right)^{-1}$ we deal with the
lagrangian proposed by Zim\'anyi and Moszkowski~\cite{kn:zimmos}. (In the
following, if not expressed explicitly otherwise, we use units such that
$\ \hbar = c = 1$.)

The field equations are
%---------------------------------------------------------------------------
\begin{equation}\label{Dirac}
\Big\{(i \partial_\alpha - g_{\rm v}  A_\alpha(x)) -
m^*\left( \phi(x)\right)\Big\}\ \psi(x) = 0
\end{equation}
%---------------------------------------------------------------------------
for the nucleons,
%---------------------------------------------------------------------------
\begin{equation}\label{Wavephi}
(\partial^\beta \partial_\beta + \mu^2_{\rm s})\ \phi(x) =
-\ {d \over{ d \phi }} m^* \left( \phi(x) \right) \
\bar{\psi}(x)\psi(x)
\end{equation}
for the scalar field and
%---------------------------------------------------------------------------
\begin{equation}\label{WaveA}
(\partial^\beta \partial_\beta + \mu_{\rm v}^2)\ A^\alpha(x) =
g_{\rm v} \,  \bar{\psi}(x) \gamma^\alpha \psi(x)
\end{equation}
%---------------------------------------------------------------------------
for the vector field.

In the mean-field approximation the source terms are replaced
by their expectation values
$\bar{\psi}(x)\psi(x) \Rightarrow \rho_{\rm s}(x)$
and
$\bar{\psi}(x) \gamma^\alpha \psi(x) \Rightarrow j^\alpha(x)$
so that the scalar and vector fields
become classical fields. For classical fields, however,
an arbitrarily weak time dependence in the scalar density
$\rho_{\rm s}(x)$ or current density $j^\alpha(x)$ results in a radiation
field which travels away from the nucleons. The total field energy
in the radiation may even be less than the masses of the field quanta.
Since the fields are only effective fields, and for the scalar field there
is no corresponding elementary particle, the physical meaning of this
radiation is obscure.

As we are using the relativistic scalar and vector fields only
for the mean-field part of the nuclear interaction we follow the
suggestion
\cite{Fel1,Fel2} to exclude radiation of fictitious mesons
from the very beginning by
using the action-at-a-distance formulation with the
symmetric Green's function of Wheeler and Feynman \cite{Wheeler}.
%---------------------------------------------------------------------------
\begin{equation}
G_{\rm s,v}(x-y) = \frac{1}{2}
 \Big( G^{advanced}_{\rm s,v} (x-y)+  G^{retarded}_{\rm s,v} (x-y)\Big) \
\end{equation}
%---------------------------------------------------------------------------
The formal solution of the wave Eq.~(\ref{Wavephi})
(for $m^* = m - g_{\rm s} \phi$)
%---------------------------------------------------------------------------
\begin{equation}\label{Phigreens}
\phi(x) = g_{\rm s} \int\! d^4 y \
 G_{\rm s} (x-y) \bar{\psi}(y)\psi(y)
\end{equation}
%---------------------------------------------------------------------------
and of Eq. (\ref{WaveA})
%---------------------------------------------------------------------------
\begin{equation}\label{Agreens}
A^\alpha(x) = g_{\rm v} \int\! d^4 y \
 G_{\rm v} (x-y) \bar{\psi}(y)\gamma^\alpha\psi(y)
\end{equation}
%---------------------------------------------------------------------------
fulfills the desired boundary condition of vanishing incoming and outgoing
free fields.

Eliminating the fields from the lagrangian (\ref{Lagr1}) leads to the
non-local action which contains only nucleon variables:
%---------------------------------------------------------------------------
\begin{eqnarray}\label{Action}
\int d^4x\ {\cal L}(x) &  = &
\int d^4x\ \bar{\psi}(x) (\gamma^\alpha i\partial_\alpha - M)\psi(x)
\nonumber \\
&+& \frac{1}{2}g_{\rm s}^2 \int d^4x\ d^4y\ \bar{\psi}(x)\psi(x)\,
G_{\rm s}(x-y)\,
 \bar{\psi}(y)\psi(y) \nonumber \\
&-& \frac{1}{2}g_{\rm v}^2 \int d^4x\ d^4y\ \bar{\psi}(x)\gamma^\alpha\psi(x)
  \,G_{\rm v}(x-y)\, \bar{\psi}(y)\gamma_\alpha\psi(y)
\end{eqnarray}
%---------------------------------------------------------------------------
If quantum effects are negligible the scalar density and vector current density
can be represented in terms of the world lines of the nucleons as

%---------------------------------------------------------------------------
\begin{eqnarray}\label{rhos}
\rho_{\rm s}(x):= \bar{\psi}(x)\psi(x) &\rightarrow&
 \sum_{\rm j=1}^{\rm A}
\int\! d \tauj \ \delta^4 (x-r_{\rm j}(\tauj))  \\
j^\mu(x):= \bar{\psi}(x)\gamma^\mu\psi(x) &\rightarrow&
 \sum_{\rm j=1}^{\rm A}
\int\! d \tauj \ \delta^4 (x-r_{\rm j}(\tauj))\,
u^\mu_{\rm j}(\tauj)\ ,
\label{rhos2}
\end{eqnarray}
%---------------------------------------------------------------------------
where $r_{\rm j}(\tauj)$ denotes the world line of the nucleon
$j$ at its proper time $\tauj$ and $u_{\rm j}(\tauj)$ its
4-velocity. Nucleons in nuclei can however not be localized well enough for
treating them as classical particles on world lines. The Fermi momentum sets a
limit to the radius of the nucleon wave packet in coordinate space which is of
the order of 2 fm. These effects, the Pauli principle and other quantum
effects, are properly taken care of in Antisymmetrized Molecular Dynamics
(AMD)~\cite{kn:amd} and in Fermionic Molecular Dynamics (FMD)~\cite{kn:fmd}.
However, presently neither the AMD nor the FMD equations of motion can be
solved numerically for large systems like gold on gold. Therefore, we mimic the
finite size effect of the wave packets by folding the finally obtained forces
with gaussian density distributions for the nucleons. This "smearing" provides
also the prescription how to eliminate strong forces at short distances which
are taken care of by the random forces of the collision term.

Therefore, the forces to be derived in the following are only meant to
represent the long range part of the nuclear interactions which constitute a
mean field.
In this sense $r_{\rm j}(\tauj)$ is the mean world line of the nucleon and
$u_{\rm j}(\tauj)$ its mean 4-velocity.

Using the representation (\ref{rhos})
the scalar field at a space-time point
$x$, for example is given by integrals over past and future proper
times $\tauj$ as
%---------------------------------------------------------------------------
\begin{equation}\label{Phiworld}
\phi(x) = g_{\rm s} \sum_{\rm j=1}^{\rm A}
        \int\! d\tauj \ G_{\rm s}(x-r_{\rm j}(\tauj)) \ .
\end{equation}
%---------------------------------------------------------------------------

By identifying the following expressions with their
representation in terms of mean positions and mean momenta of the
nucleons
%---------------------------------------------------------------------------
\begin{eqnarray}
 \int d^4x\ d^4y\ \bar{\psi}(x)\psi(x)\,G_{\rm s}(x-y)\,
 \bar{\psi}(y)\psi(y) &\rightarrow& \nonumber \\
&& \mbox{\hspace*{-6cm}}
\sum^{\rm A }_{\stackrel{{\scriptstyle \rm i,j=1}}{\rm i\neq j}}
\int\! d\tau_i d\tauj\ G_{\rm s}(r_{\rm i}(\taui)-
    r_{\rm j}(\tauj)) \quad ,\\
 \int d^4x\ d^4y\ \bar{\psi}(x)\gamma^\alpha\psi(x)
  \,G_{\rm v}(x-y)\, \bar{\psi}(y)\gamma_\alpha\psi(y)
 &\rightarrow& \nonumber \\
&&\mbox{\hspace*{-6cm}}
 \sum^{\rm A}_{\stackrel{{\scriptstyle \rm i,j=1}}{\rm i\neq j}}
\int\! d\taui d\tauj\ G_{\rm v}(r_{\rm i}(\taui)-
   r_{\rm j}(\tauj))\,
u_{{\rm i}\alpha}(\taui) u_{\rm j}^\alpha(\tauj)
\end{eqnarray}
%---------------------------------------------------------------------------
in the action
(\ref{Action}) one obtains the non-instantaneous
action-at-a-distance \cite{Wheeler}
%--------------------------------------------------------------------------
\begin{eqnarray}\label{Action1}
\int\! d^4x \ {\cal L}(x) &\rightarrow&  \nonumber \\
{\cal A} &=& - \frac{1}{2} \sum^{\rm A}_{\rm i=1} \int\! d\taui
\left(M-\lambda_{\rm i}(\taui)\right) u_{\rm i}(\taui)^2
  \nonumber \\
&+&\frac{1}{2} g^2_{\rm s}
\sum^{\rm A}_{\stackrel{{\scriptstyle \rm i,j=1}}{\rm i\neq j}}
\int\! d\taui d\tauj \ G_{\rm s}(r_{\rm i}(\taui) -
  r_{\rm j}(\tauj)) \\
&-&\frac{1}{2} g^{2}_{\rm v}
\sum^{\rm A}_{\stackrel{{\scriptstyle \rm i,j=1}}{\rm i\neq j}}
\int\! d\taui d\tauj \ G_{\rm v}(r_{\rm i}(\taui) -
  r_{\rm j}(\tauj))\
    u_{{\rm i}\alpha}(\taui) u_{\rm j}^\alpha(\tauj) \ .\nonumber
\end{eqnarray}
%---------------------------------------------------------------------------
Lagrange multipliers $\lambda_{\rm i}(\taui)$ are introduced
to ensure $u_{\rm i}(\taui)^2=1$.

The equations of motion which result from this Wheeler-Feynman action
are equivalent to the classical field equations, provided
the system does not radiate \cite{Wheeler}.

However, the Wheeler Feynman equations of motion cannot be
solved in general
as one needs to know the world lines for all the
past and future times in order to calculate the fields which enter
the equations for the world lines~\cite{BelMartin}. Furthermore, the
no-interaction theorem states that, except in the case of
non-interacting particles, there exist no covariant
equations of motion for world lines in which only the 4-positions and
the 4-velocities at a given time enter, as is the case in
non-relativistic mechanics.

In the following this no-interaction theorem is circumvented by
introducing an approximative solution to the non-local Wheeler Feynman
equations of motion. This is achieved by the so called
small acceleration approximation which does not assume small
velocities.

\subsection*{\large\sl 2.1. Action-at-a-distance in the small acceleration
approximation}

In order to conserve manifest covariance of the equations of motion
we introduce first the concept of a scalar time. For that the
whole Minkowski space is chronologically ordered by a set of
space like surfaces $S(x)$ which attribute to each 4-position
$x$ a scalar time $t$ by
%---------------------------------------------------------------------------
\begin{equation}\label{Isochrone}
S(x)=t \ \ , \ \ \  \mbox{with} \ \ \ \partial^0 S(x) > 0 \ ,\
\partial^0 \partial^0 S(x) \ge 0 \ .
\end{equation}
%---------------------------------------------------------------------------
The simplest choice for these isochrones, which we shall take
in the following, is
$S(x)=\eta_\alpha x^a$, where $\eta_\alpha$ is a position
independent time like 4-vector.

Unlike in a collision the nucleons are not strongly accelerated by the
action of the mean-field. Therefore, in order to describe the motion
in the mean-field one may expand each world line
around the proper time $\tau^s_{\rm j}$ at which the particle is at
a given isochrone, i.e. $t=S(r_{\rm j}(\tau^s_{\rm j}))$ for each $j$.

%---------------------------------------------------------------------------
\begin{equation}\label{worldline}
r_{\rm j}^\alpha(\tauj)=r_{\rm j}^\alpha(\tau^s_{\rm j})
  +(\tauj-\tau^s_{\rm j})
u_{\rm j}^\alpha(\tau^s_{\rm j})+
\frac{1}{2}(\tauj-\tau^s_{\rm j})^2 \,
a_{\rm j}^\alpha(\tau^s_{\rm j})+\cdots \ .
\end{equation}
%---------------------------------------------------------------------------
This way we are defining a synchronization prescription for all particles,
which does not
depend on the frame. For calculating the fields we neglect of the
quadratic term with the acceleration
$a_{\rm j}^\alpha(\tau^s_{\rm j})$ and all higher powers
which leaves us with a straight world line in the vicinity of the
synchronizing time $t=S(r_{\rm j}(\tau^s_{\rm j}))$ (c.f. Fig.~1).
This is called ''small acceleration approximation"
\begin{figure}[t]
\vspace*{-10mm}
\insertplot{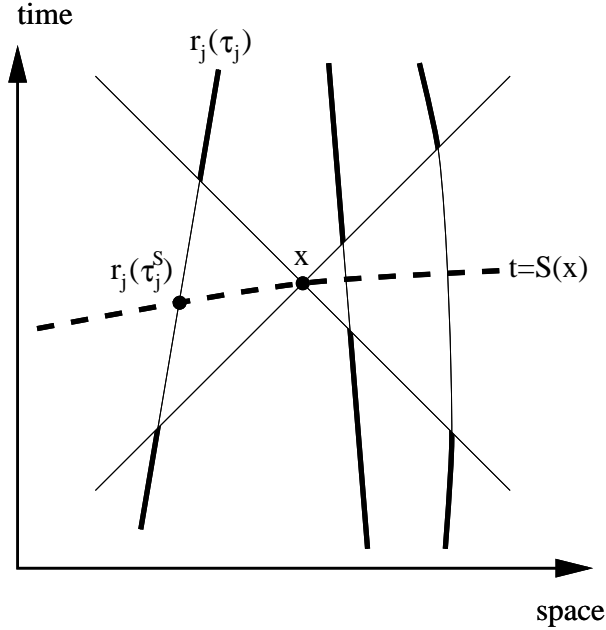}
\vspace*{-10mm}
\caption{World lines and synchronizing hypersurface $S(x)$}
\end{figure}

The small acceleration approximation is of course best in the vicinity
of $r_{\rm j}^\alpha(\tau^s_{\rm j})$ and becomes worse further away.
As sketched in Fig. 1 only those parts of the world lines (thick lines),
which are outside the light cone (centered at $x$),
contribute to the field strength at point $x$.
A world line which hits the light cone far away from $x$ may be badly
approximated by Eq. (\ref{worldline}), but for short range interactions
a distant particle does not contribute anymore to the field at $x$.

Thus, the first condition for the validity of the approximation is
that the range $\mu^{-1}$ is small compared to the curvature of the
world lines, i.e. the inverse of the acceleration.
The second condition is weak radiation, which is fulfilled when the
acceleration is small compared to the meson mass $\mu$. Both
conditions are actually the same, namely
%---------------------------------------------------------------------------
\begin{equation}\label{smallacc}
\mid a_\mu a^\mu \mid \ \ll \ \mu^2 .
\end{equation}
%---------------------------------------------------------------------------

The assumption of small accelerations is justified if the $\phi$
and $A^\alpha$ fields are only meant to be the mean-field part of the
nucleon-nucleon interaction in a hadronic surrounding. The hard collisions
between individual nucleons which are due to the repulsive core will
cause large accelerations and also create new particles.
These hard collisions cannot be described within Walecka type
mean-field models.
Therefore, it is consistent to regard $\phi (x)$ and $A^\alpha(x)$ as Hartree
mean-fields which bring about only small accelerations and which
are not radiated away from their sources.

Inserting the straight world line into Eq.
(\ref{Phiworld})
results in the easily understood situation that the field at a point $x$ is
just the sum of Lorentz-boosted Yukawa potentials which are traveling
along with the charges:
%---------------------------------------------------------------------------
\begin{equation}\label{Phiyuk}
\phi(x)  =  \frac{g_{\rm s}}{4\pi}  \sum_{\rm j}
\frac{\exp\left\{-\mu_{\rm s} \sqrt{R_{\rm j}(x)^2}\right\} }
{\sqrt{R_{\rm j}(x)^2}}\ ,
\end{equation}
%---------------------------------------------------------------------------
where $R_{\rm j}(x)^2$ is given by
%---------------------------------------------------------------------------
\begin{equation}
R_{\rm j}(x)^2 = - (x-r_{\rm j}(\tau^s_{\rm j}))^2 +
\left[(x^\alpha-r_{\rm j}^\alpha(\tau^s_{\rm j}))\,
u_{{\rm j}\alpha}(\tau^s_{\rm j})\right]^2
\end{equation}
%---------------------------------------------------------------------------
The vector field is derived in an analogue fashion as
%---------------------------------------------------------------------------
\begin{equation}\label{Ayuk}
A^\alpha(x) = \frac{g_{\rm v}}{4\pi} \sum_{\rm j}
\frac{\exp\left\{-\mu_{\rm v}\sqrt{R_{\rm j}(x)^2}\right\} }
{\sqrt{R_{\rm j}(x)^2}}\, u_{\rm j}^\alpha(\tau^s_{\rm j}) \ .
\end{equation}
%---------------------------------------------------------------------------
At this level of the approximation the causality problem with the advanced
part of the Green function is not present because the retarded and the
advanced fields are identically the same when they are
created by charges moving on straight world lines.
Therefore one may regard the fields as retarded only.
In addition the unsolved problem of radiation reaction, where
the radiation acts back on the world lines \cite{Jackson},
does not occur because there are no radiation fields any more.

\subsection*{\large\sl 2.2. Instantaneous action-at-a-distance}

In the spirit of the small acceleration assumption discussed in the previous
section one can use the straight line expansion in the action
(\ref{Action1}) and perform the integration over
$\tauj$. This results in an instantaneous
action-at-a-distance where the Lorentz-boosted Yukawa fields appear again
and there is only one time, the  scalar synchronizing
time $t$.
%---------------------------------------------------------------------------
\begin{eqnarray}\label{Action2}
{\cal A} = \int \! dt \sum_{\rm i} \frac{1}{\partial_\alpha
  S(r_{\rm i}(t))u_{\rm i}^\alpha(t)}
 \Bigg[
&-&\frac{1}{2} \left(M-\lambda_{\rm i}(t)\right)u_{\rm i}(t)^2  \nonumber\\
&+& \frac{1}{2} \sum_{\stackrel{\scriptstyle\rm j}{\rm j \neq i}}
\frac{g^2_{\rm s}}{4\pi}
\frac{\exp\{-\mu_{\rm s} \sqrt{R_{\rm ij}(t)^2}\}}{\sqrt{R_{\rm ij}(t)^2}} \\
&& \mbox{\hspace*{-2cm}}-
\frac{1}{2} \sum_{\stackrel{\scriptstyle\rm j}{\rm j \neq i}}
\frac{g^2_{\rm v}}{4\pi}
\frac{\exp\{-\mu_{\rm v} \sqrt{R_{\rm ij}(t)^2}\}}{\sqrt{R_{\rm ij}(t)^2}}\,
u_{{\rm i}\alpha}(t) u_{\rm j}^\alpha(t)\,\Bigg]
\nonumber
\end{eqnarray}
%---------------------------------------------------------------------------
The four-positions $r_{\rm i}^\alpha(t)\equiv
r_{\rm i}^\alpha(\taui(t))$ and
4-velocities $u_{\rm i}^\alpha(t) \equiv u_{\rm i}^\alpha(\taui(t))$
are to be taken at the same scalar synchronizing time $t$ and
%---------------------------------------------------------------------------
\begin{equation}
R_{\rm ij}^2(t) := - (r_{\rm i}(t) - r_{\rm j}(t))^2 +
\Big[(r_{{\rm i}\beta}(t) -
r_{{\rm j}\beta}(t)) u_{\rm j}^\beta (t)\Big]^2
\end{equation}
%---------------------------------------------------------------------------

The small acceleration approximation together with the introduction of a
synchronizing hypersurface $S(x)$ leads to an equal time lagrangian which is
Lorentz-scalar and written in a manifestly covariant way.

Giving up explicit covariance and choosing $\eta=(1,0,0,0)$ in a special
coordinate frame, the positions and velocities take the form
%---------------------------------------------------------------------------
\begin{equation}\label{eq:rcm}
r_{\rm i}(\taui(t)) = \Big(t\, , \,\vec{r}_{\rm i}(t)\Big)
\ \ \ \mbox{and} \ \ \ u_{\rm i}(\taui(t))
= \frac{1}{\sqrt{1-\vec{v}_{\rm i}^{\,2}(t)}} \Big( 1\, ,\,
  \vec{v}_{\rm i}(t) \Big) \ .
\end{equation}
%---------------------------------------------------------------------------
With that a Lagrange function ${\cal L} (\vec{r}_{\rm i}(t),
\vec{v}_{\rm i}(t))$
can be defined which depends only on the independent variables and one time.
Even the Lagrange multipliers $\lambda_{\rm i}(t)$ are not needed anymore if
the
variation is with respect to $\vec{v}_{\rm i}(t)$ instead of all four
$u_{\rm i}^\alpha(t)$.

The advantage of the instantaneous Lagrangian is that one can
define easily the hamiltonian and the total momentum, which are then strictly
conserved by the equations of motion.

\subsection*{\large\sl 2.3. Hamilton equations of motion}

In the following we want to express the total hamiltonian as a function of the
positions $\vec{r}_{\rm i}$ and the canonical momenta $\vec{p}_{\rm i}$.
Using relations
(\ref{Action2})-(\ref{eq:rcm}) the Lagrange function is
\begin{eqnarray}
\label{eq:lang}
{\cal L} = &-&\sum_{\rm i} \ \left[ {m \over{ u_{\rm i}^0 }} + {1 \over 2}
\sum_{\stackrel{\scriptstyle\rm j}{\rm j \neq i}}\
{g^2_{\rm s} \over{ 4\pi u_{\rm i}^0 }}
{{\rm exp}\left\{ -\mu_{\rm s} \sqrt{R_{\rm ij}^2(t)} \right\} \over{
\sqrt{R_{\rm ij}^2(t)} }} \right. \\
&-& \left. {1 \over 2}
\sum_{\stackrel{\scriptstyle\rm j}{\rm j \neq i}}\
{g^2_{\rm v} \over{ 4\pi
u_{\rm i}^0 }}
{{\rm exp}\left\{ -\mu_{\rm v} \sqrt{R_{\rm ij}^2(t)} \right\} \over{
\sqrt{R_{\rm ij}^2(t)} }} u_{\rm i\alpha}(t) u_{\rm j}^{\alpha}(t)
  \right] \, . \nonumber
\end{eqnarray}
The canonical momenta and the energy are determined from the Lagrange function
above as
\begin{eqnarray}
\vec{p}_{\rm i} &=& {\partial {\cal L} \over{ \partial \vec{v}_{\rm i} }}
\quad , \\
E &=& \sum_{\rm i} {\partial {\cal L} \over{ \partial \vec{v}_{\rm i} }}
\vec{v}_{\rm i} - \cal{L} \quad .\nonumber
\end{eqnarray}
Performing these derivatives, the energy is given by
\begin{eqnarray}
\label{eq:enu}
E &=& \sum _{\rm i} \ m\ u_{\rm i}^0 - {1\over 2}
\sum_{\stackrel{\scriptstyle\rm i,j}{\rm j \neq i}}\
  f_{\rm ij} u_{\rm i}^0 - {1\over 2}
\sum_{\stackrel{\scriptstyle\rm i,j}{\rm j \neq i}}\
  \widehat{f}_{\rm ji}\ (\vec{r}_{\rm ij} \vec{u}_{\rm i})^2
{\left( u_{\rm i}^0 \right)^2 \over{ u_{\rm j}^0 }} \nonumber \\
&+& {1\over 2}
\sum_{\stackrel{\scriptstyle\rm i,j}{\rm j \neq i}}\
  g_{\rm ji}\ u_{\rm i}^0 - {1\over 2}
\sum_{\stackrel{\scriptstyle\rm i,j}{\rm j \neq i}}\
  g_{\rm ji} \left[
u_{\rm i}^0 \vec{u}_{\rm i}^2 - {\left( u_{\rm i}^0 \right)^2
\over{ u_{\rm j}^0 }} \vec{u}_{\rm i} \vec{u}_{\rm j} \right] \\
&+& {1\over 2}
\sum_{\stackrel{\scriptstyle\rm i,j}{\rm j \neq i}}\
  \widehat{g}_{\rm ji}\ (\vec{r}_{\rm ij} \vec{u}_{\rm i})^2 \left[
\left( u_{\rm i}^0 \right)^3 - {\left( u_{\rm i}^0 \right)^2
 \over{ u_{\rm j}^0 }}
\vec{u}_{\rm i} \vec{u}_{\rm j} \right] , \nonumber
\end{eqnarray}

\noindent where we use the abbreviations
\begin{eqnarray}
\label{eq:fdef1}
f_{\rm ij} \equiv f(R_{\rm ij}) = {g_{\rm s}^2\over{4 \pi}}\ {{\rm exp}\left\{
-\mu_{\rm s} \rm R_{ij}\right\} \over{R_{\rm ij}}} \quad &,& \quad
g_{\rm ij} = {g_{\rm v}^2\over{4 \pi}}\ {{\rm exp}\left\{ -\mu_{\rm v}
\rm R_{ij}\right\} \over{R_{\rm ij}}} \, , \nonumber \\
\widehat{f}_{\rm ji} \equiv -{1 \over{ R_{\rm ji} }} {d \over{ d R_{\rm ji} }}
f(R_{\rm ji}) =  {g_{\rm s}^2\over{4 \pi}}\ (1 &\hspace*{-5mm}+& \hspace*{-5mm}
  \mu_{\rm s} R_{\rm ji})
{{\rm exp}\left\{ -\mu_{\rm s} \rm R_{ji}\right\} \over{ R_{\rm ji}^3 }}\, , \\
R_{\rm ij}^2 = \vec{r}_{\rm ij}^2 + (\vec{r}_{\rm ij}
\vec{u}_{\rm j})^2 \quad &,& \quad
  \vec{r}_{\rm ij} = \vec{r}_{\rm i} - \vec{r}_{\rm j} \, ,
\label{eq:fdef2}
\end{eqnarray}
and $\widehat{g}_{\rm ji}$ is defined in the same way as $\widehat{f}_{\rm
ji}$.
It is worthwhile to note that $f_{\rm ij} \neq f_{\rm ji}$.

The momentum of a particle looks as
\begin{eqnarray}
\label{eq:mom}
\vec{p}_{\rm i} &=& m\vec{u}_{\rm i} - {1\over 2} \left(
\sum_{\stackrel{\scriptstyle\rm j}{\rm j \neq i}}\
  f_{\rm ij} \right) \vec{u}_{\rm i} - {1\over 2}
\sum_{\stackrel{\scriptstyle\rm j}{\rm j \neq i}}\
  \widehat{f}_{\rm ji} {u_{\rm i}^0 \over{ u_{\rm j}^0 }} \left[
\vec{r}_{\rm ij} (\vec{r}_{\rm ij} \vec{u}_{\rm i})
+ \vec{u}_{\rm i} (\vec{r}_{\rm ij}
\vec{u}_{\rm i})^2 \right] \nonumber \\
&+&  {1\over 2}
\sum_{\stackrel{\scriptstyle\rm j}{\rm j \neq i}}\
  \left( g_{\rm ij} + {u_{\rm i}^0
\over{ u_{\rm j}^0 }} g_{\rm ji} \right) \vec{u}_{\rm j}
- {1\over 2}
\sum_{\stackrel{\scriptstyle\rm j}{\rm j \neq i}}\
  g_{\rm ji} \left[
\left( u_{\rm i}^0 \right)^2 - {u_{\rm i}^0 \over{ u_{\rm j}^0 }}
  \vec{u}_{\rm i} \vec{u}_{\rm j} \right] \vec{u}_{\rm i} \\
&+& {1\over 2}
\sum_{\stackrel{\scriptstyle\rm j}{\rm j \neq i}}\
  \widehat{g}_{\rm ji} \left[
\left( u_{\rm i}^0 \right)^2 - {u_{\rm i}^0 \over{ u_{\rm j}^0 }}
  \vec{u}_{\rm i} \vec{u}_{\rm j} \right]
 \left[
  \vec{r}_{\rm ij} (\vec{r}_{\rm ij} \vec{u}_{\rm i}) +
  \vec{u}_{\rm i} (\vec{r}_{\rm ij} \vec{u}_{\rm i})^2
 \right] \, . \nonumber
\end{eqnarray}

Let us first prove that for isotropic nuclear matter, like in the original
mean-field picture~\cite{kn:serwal}
the vector potential will not contribute to the canonical momentum.
In that case the summation can be replaced by
an integration over space and momentum, folded with the phase space
distribution
function. Since for isotropic nuclear matter the distribution function is
independent of the position, the space integral can be carried out
immediately and the remaining part is written as a summation over momenta.
This way we get for the Yukawa terms in nuclear matter
\begin{eqnarray}
\sum_{\rm j}\  f_{\rm ij} &\rightarrow&
  \sum_{\rm j}\
  \int d^3r_{\rm j} \ f(R_{\rm ij}) = \left( {g_{\rm s} \over{
   \mu_{\rm s}}} \right)^2 \sum_{\rm j}\
  {1\over{u_{\rm j}^0}} \nonumber \, . \\
\sum_{\rm j}\ \widehat{f}_{\rm ji}\ (\vec{r}_{\rm ij} \vec{u}_{\rm i})^2
&\rightarrow&
  \sum_{\rm j}\
\int d^3r_{\rm j}\ \widehat{f}(R_{\rm ji}) (\vec{r}_{\rm ij}
\vec{u}_{\rm i})^2 =
  \left( {g_{\rm s} \over{ \mu_{\rm s}}} \right)^2
    {\vec{u}_{\rm i}^2\over{\left( u_{\rm i}^0 \right)^3 }} \
    \sum_{\rm j}\  1 \, .
\label{eq:nms}
\end{eqnarray}

Substituting the above expressions into the momentum~(\ref{eq:mom}) the
$\widehat{f}_{\rm ij}$ term gives the same contribution as the
$f_{\rm ij}$, while
the $g_{\rm ij}$ terms cancel each other and the momentum can be written as

\begin{equation}
\vec{p}_{\rm i} = m \vec{u}_{\rm i} - \left( {g_{\rm s}\over{
\mu_{\rm s} }} \right)^2 \vec{u}_{\rm i}
  \sum_{\stackrel{{\scriptstyle\rm j}}{\rm j\neq i}} \
{1\over{u_{\rm j}^0}} + \left( {g_{\rm v}\over{\mu_{\rm v} }}
\right)^2
  \sum_{\stackrel{{\scriptstyle\rm j}}{\rm j\neq i}} \
{\vec{u}_{\rm j}\over{ u_{\rm j}^0}} \, .
\label{eq:pnm0}
\end{equation}

The last term gives zero, since the average of $\vec{u}_{\rm j}$ is zero in
nuclear matter and Eq.~(\ref{eq:pnm0}) can be written as

\begin{equation}
\vec{p}_{\rm i} = m^* \vec{u}_{\rm i} \quad ,
\label{eq:pnm1}
\end{equation}

\noindent where

\begin{eqnarray}
m^* &=& m - \left( {g_{\rm s}\over{\mu_{\rm s} }} \right)^2
  \sum_{\stackrel{{\scriptstyle\rm j}}{\rm j\neq i}} \
{1\over{u_{\rm j}^0}} \nonumber \\
&=& m - \left( {g_{\rm s}\over{\mu_{\rm s} }} \right)^2
\sum_{\rm j} \ {m^* \over{\sqrt{{m^*}^2 + \vec{p}_{\rm j}^2}}} \, .
\label{eq:msnm}
\end{eqnarray}
Similarly, the energy density (\ref{eq:enu}) has the form
\begin{equation}
\epsilon = \sum_{\rm i}\ \sqrt{{m^*}^2 + \vec{p}_{\rm i}^2}
+ {1\over 2} \left( {g_{\rm v}\over{\mu_{\rm v} }} \right)^2
\left( j^0 \right)^2 + {1\over 2} \left( {\mu_{\rm s}
 \over{ g_{\rm s} }} \right)^2
(m - m^*)^2 \quad .
\label{eq:ednm}
\end{equation}

\noindent Eqs.~(\ref{eq:msnm}-\ref{eq:ednm}) are just
the mean-field results of the Walecka model~\cite{kn:serwal}.

It is worthwhile to mention that the small acceleration approximation which
introduces $R_{\rm ij}$ instead of $\mid \vec{r}_{\rm ij}\mid$ in
Eqs.~(\ref{eq:fdef1}-\ref{eq:fdef2}), is
definitely needed to get back the relativistic mean-field result for nuclear
matter.
Therefore, for the coupling strengths $g_{\rm s}$ and $g_{\rm v}$ we can simply
take the values obtained from the saturation properties of the nuclear
matter~\cite{Lindner}.

For the Hamilton equations we need the energy as the function of the
canonical momenta instead of the 4-velocities $u_{\rm i}$. For that we have
to invert
the $p(u)$ equation (\ref{eq:mom}) into a $u(p)$ one. One can do this
approximatively by first expanding the 4-velocities in the $p(u)$ relation
up to the third order in the quantities $(\vec{p}/m)^2$, $f_{\rm ij}$ and
$g_{\rm ij}$, and then substituting that expression for $u(p)$ into the
energy Eq.~(\ref{eq:enu}). At the end we will
check the validity of this approximation. After tedious calculations
the energy can be expressed as follows (in the following we include explicitly
$c$ and $\hbar$)

\begin{eqnarray}
\label{eq:en}
{E\over{mc^2}} &=& \sum_{\rm i}\ \sqrt{1+\vec{\tilde{p}}_{\rm i}^2}
- {1\over 2}
\sum_{\stackrel{{\scriptstyle \rm i,j}}{\rm j \neq i}}\
{\tilde{f}_{\rm ij} \over{
\sqrt{1+\vec{\tilde{p}}_{\rm i}^2}}} + {1\over 2}
\sum_{\stackrel{{\scriptstyle \rm i,j}}{\rm j \neq i}}\
\tilde{g}_{\rm ji} \sqrt{1+\vec{\tilde{p}}_{\rm i}^2} \nonumber \\
&-& {1\over 2}
\sum_{\stackrel{{\scriptstyle \rm i,j}}{\rm j \neq i}}\
\tilde{g}_{\rm ji}
{(\vec{\tilde{p}}_{\rm i} \vec{\tilde{p}}_{\rm j}) \over{
\sqrt{1+\vec{\tilde{p}}_{\rm i}^2}}} + {1\over 2}
\sum_{\rm i}\ \left(
\sum_{\stackrel{{\scriptstyle \rm j}}{\rm j \neq i}}\
  \tilde{f}_{\rm ji} \vec{\tilde{p}}_{\rm i}
- \sum_{\stackrel{{\scriptstyle \rm j}}{\rm j \neq i}}\
  \tilde{g}_{\rm ji} \vec{\tilde{p}}_{\rm j}
\right)^2 \quad ,
\end{eqnarray}

\noindent where
\begin{eqnarray}
\tilde{f}_{\rm ij} &=& {1\over{4\pi}} \left( {g_{\rm s} m \over{
\mu_{\rm s}}} \right)^2 \left( {\hbar \over{ mc }} \right)
{{\rm exp}\left\{ -{\hbar \over{ \mu_{\rm s} c }}
  \tilde{R}_{\rm ij}\right\} \over{\tilde{R}_{\rm ij}}} \quad {\rm and}
\nonumber \\
\tilde{R}_{\rm ij}^2 &=& \vec{r}_{\rm ij}^2 + ( \vec{r}_{\rm ij}
\vec{\tilde{p}}_{\rm j} )^2 \quad ,
\end{eqnarray}

\noindent with $\vec{\tilde{p}}_{\rm i} = \vec{p}_{\rm i}/mc$.
$\tilde{g}_{\rm ij}$ is defined similarly to $\tilde{f}_{\rm ij}$. The last
term is already of third order and hence small.

{}From Eq.~(\ref{eq:en}) the Hamilton equations
\begin{equation}
{d \over{ dt }} \vec{r}_{\rm i} = -\
{\partial E\over{\partial \vec{p}_{\rm i}}} \qquad
{d \over{ dt }} \vec{p}_{\rm i} =
  {\partial E\over{\partial \vec{r}_{\rm i}}}
\end{equation}
are calculated easily.

\section*{\large\bf 3. Details of the QMD calculations}

Molecular dynamics is a classical many-body theory in which some quantum
features
due to the fermionic nature of nucleons are simulated.  For the details of
the theory we would like to refer the reader to the works of
Aichelin~\cite{kn:aich} and the
Frankfurt group~\cite{kn:frank}. In our calculation we introduced some
modifications which are described in~\cite{kn:judqmd,kn:paula}.
In the following this
model is used in a relativistic treatment with scalar and vector forces.

\subsection*{\large\sl 3.1. The mean-field forces}
The scalar and vector forces are the relativistic two-body forces derived in
the previous section which give saturation for
nuclear matter even without an explicit density dependent term. However,
Walecka's values for the coupling constants result in a too
high compressibility. For this reason it seems to be more adequate to use
instead the Zim\'anyi-Moszkowski (ZM) lagrangian~\cite{kn:zimmos} which
provides a more reasonable compressibility. The different coupling
of the mean-field ($m^* = ( m + g_{\rm s} \phi )^{-1}$)
modifies the scalar part of the total energy.
In the used approximation, the expression

\begin{equation}
-{1\over 2} \sum_{\stackrel{\scriptstyle \rm i,j}{
\scriptstyle \rm j \neq i}} \
{\tilde{f}_{\rm ij}
 \over{
\sqrt{1 + \vec{\tilde{p}}_{\rm i}^2} }}
  \left( {1 \over{
1 + 2 {\displaystyle \sum_{\stackrel{\scriptstyle \rm k}{\rm k \neq j}}\
\tilde{f}_{\rm jk} } }} \right) \quad ,
\end{equation}
replaces the second term on the r.h.s. of Eq.~(\ref{eq:en}). Here we see the
effect of the nonlinear coupling in the ZM lagrangian. The strength of the
scalar potential which is felt by particle j is weakened.

We compare the Walecka forces with the ZM ones and study
their effect on nuclear multifragmentation. The coupling strengths $g_{\rm s}$
and $g_{\rm v}$ are determined from the ground state properties of nuclear
matter~\cite{Lindner,kn:serwal}. In order to include the symmetry
energy we take also
the $\varrho$ meson into account. To
simplify the calculations the mass of the $\varrho$ meson is taken to be equal
to the one of the $\omega$ meson, and the coupling constants are fitted to get
the appropriate symmetry energy. In Table~1 we list all coupling
constants used.
\begin{table}[t]
\begin{center}
\begin{tabular}{||l|c|c|c|c|c||} \hline
& $g_{\rm s}$ & $g_{\rm v}$ & $g_{\varrho}$ & $\mu_{\rm s} c^2$ &
$\mu_{\rm v}  c^2$ = $\mu_{\varrho} c^2$ \\ \hline \hline
W & 11.04 & 13.74 & 7.0 & 550 MeV & 783 MeV \\ \hline
ZM & 7.84 & 6.67 & 7.0 & 550 MeV & 783 MeV \\ \hline
\end{tabular}
\end{center}
\caption{Parameters of the Walecka and the Zim\'anyi-Moszkowski forces}
\end{table}

Since the relativistic two-body forces are meant to describe the long
range part of the interaction, the Yukawa functions are folded with a gaussian
density distributions for the nucleons.
For nuclear matter one gets a smooth density distribution if the
gaussians $e^{-\alpha^2 (\vec{r} - \vec{r}_1 )^2}$ have a width parameter of
$\alpha=0.5$ fm$^{-1}$. We use that value for
folding the Yukawa forces, analogue to our earlier QMD
calculations~\cite{kn:judqmd,kn:paula}, with $R^2_{\rm ij}$ replacing
$\vec{r}^{\ 2}_{\rm ij}$ everywhere.

\subsection*{\large\sl 3.2. Initial conditions}
For the initial coordinate and momentum distribution of the particles we
use the ones, which were prepared for the non-relativistic
calculations~\cite{kn:judqmd} .
The main aspect when generating initial positions and momenta is to get a
smooth
phase-space distribution. The ground-state energies calculated with
distributions which fit nuclear charge densities
are found to be good for both, the Walecka and the ZM forces.

\subsection*{\large\sl 3.3. Cross sections}
\begin{figure}[t]
\vspace*{-10mm}
\insertplot{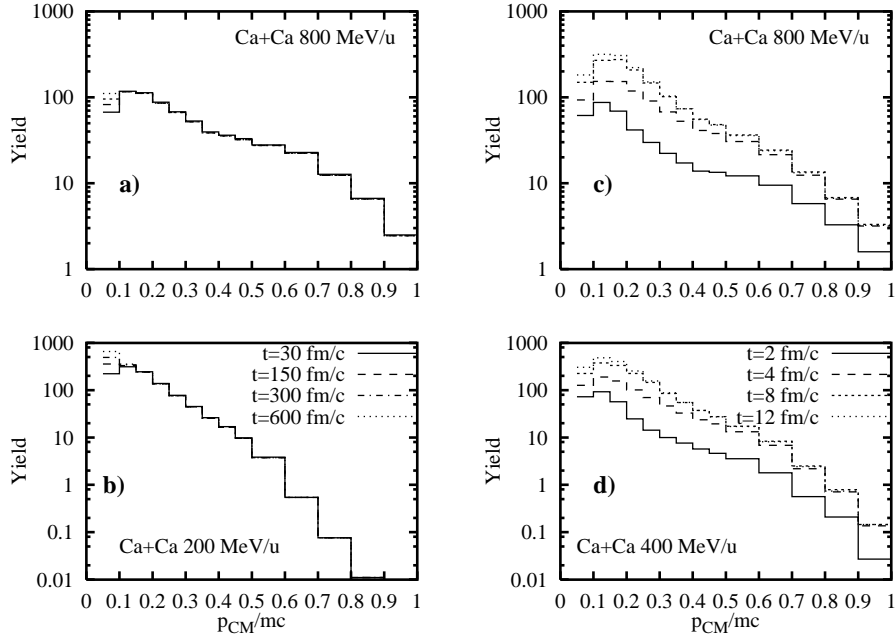}
\vspace*{-10mm}
\caption{Distribution of nucleon-nucleon relative momenta up to time $t$
in Ca+Ca central collision}
\end{figure}

The multifragmentation depends on the cross sections used. According to Cugnon,
his parameterization~\cite{kn:cug} gives a good fit only for relative
momenta $\ge 0.8$ GeV/c in the two particle center of mass frame.
For Ca + Ca central
collisions at 200, 400, 600 and 800 MeV/u initial beam energies
Fig.~2 shows however, that the vast majority of the collisions occur at
relative momenta below $0.2$ GeV/c. To see the effect of the cross section
on multifragmentation
we fit the experimental free nucleon-nucleon
cross sections~\cite{kn:land-b} as a function of the relative momentum
with second order polynomials
and use that fit in our calculations. However, due to the screening
effect of the other nucleons we do not allow collisions at
distances larger than $r_{\rm coll}$. The screening length is somewhat
arbitrary, so we work with two parameterization, $r_{\rm coll}$ = 1.6 fm
and $r_{\rm coll}$ = 2.4 fm.
The low energy cross section, we use in
the calculation, is displayed in Fig.~3. The dependence of the
fragment distribution on the different cross sections is discussed
 in the next section.
\begin{figure}
\vspace*{-10mm}
\insertplot{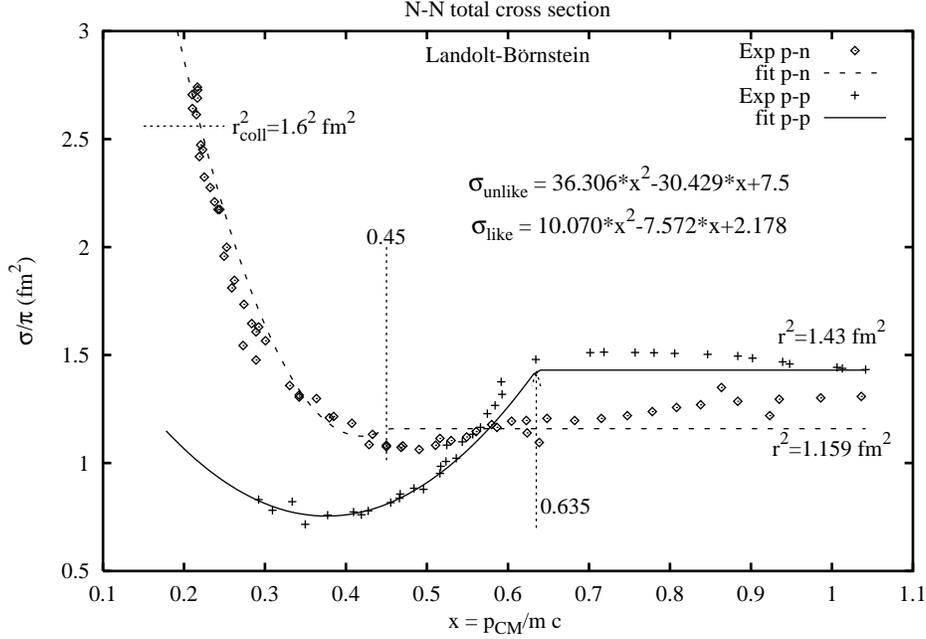}
\vspace*{-10mm}
\caption{Low energy nucleon-nucleon cross section}
\end{figure}

\subsection*{\large\sl 3.4. Effective mass in two-body collisions}

In relativistic many-body calculations single-particle energies
$e_{\rm i}$ cannot be defined such that ($e_{\rm i},\vec{p}_{\rm i}$)
form a 4-vector.
This holds true only for the total momentum $\vec{P}$ and total energy $E$.
However, we still try to deduce a single particle energy, which we use for
prescribing the conservation of energy in each collision. If one considers
only
the scalar potential in the lowest order the solution is easy. The
equation (\ref{eq:en}) in the leading order can be written as
\begin{equation}
\label{eq:efm}
{E\over{mc^2}} = \sum_{\rm i}\ \sqrt{1+\vec{\tilde{p}}_{\rm i}^2} - {1\over 2}
\sum_{\stackrel{{\scriptstyle \rm i,j}}{\rm j \neq i}}\
{\tilde{f}_{\rm ij} \over{ 1 + 2 {\displaystyle
\sum_{\stackrel{{\scriptstyle \rm k}}{\rm k \neq j}}\
\tilde{f}_{\rm jk} } }}
{1 \over{ \sqrt{1+\vec{\tilde{p}}_{\rm i}^2}}} + {1\over 2}
\sum_{\stackrel{{\scriptstyle \rm ij}}{\rm j \neq i}}\
\tilde{g}_{\rm ji} \sqrt{1+\vec{\tilde{p}}_{\rm i}^2}
\end{equation}

There are two ways in our case to define the effective mass $m^*$ appearing in
the collisions:

Case A: If we neglect the momentum dependence in $\tilde{f}_{\rm ij}$ and
$\tilde{g}_{\rm ij}$, the energy of the i$^{\rm th}$ particle can be written
as
\begin{eqnarray}
\label{eq:sinen}
{E(A) - E_{\rm i}(A-1) \over{ mc^2 }} &=& \sqrt{ 1 + \vec{\tilde{p}}_{\rm i}^2
}
- {1 \over 2}
\sum_{\stackrel{{\scriptstyle \rm j}}{\rm j \neq i}}\
  {\tilde{f}_{\rm ij} \over {
\sqrt{ 1 + \vec{\tilde{p}}_{\rm i}^2 } }}
{1 \over{ 1 + 2 {\displaystyle
  \sum_{\stackrel{\scriptstyle \rm k}{\rm k \neq j}}\
    \tilde{f}_{\rm jk} } }} \\
&+& {1 \over 2}
\sum_{\stackrel{{\scriptstyle \rm j}}{\rm j \neq i}}\
  \tilde{g}_{\rm ji} \sqrt{ 1 + \vec{\tilde{p}}_{\rm i}^2 }
+ \mbox{terms independent of }\  \vec{\tilde{p}}_{\rm i} \ ,\nonumber
\end{eqnarray}
where $E_{\rm i}(A-1)$ means the energy of the system when particle i is taken
out.

We can expand Eq.~(\ref{eq:sinen}) in $\vec{\tilde{p}}_{\rm i}^2$ and get
$$
\epsilon_{\rm i} = {1 \over 2} \vec{\tilde{p}}_{\rm i}^2 \left[ 1 + {1 \over 2}
\sum_{\stackrel{{\scriptstyle \rm j}}{\rm j \neq i}}\
  {\tilde{f}_{\rm ij} \over{ 1 + 2 {\displaystyle
  \sum_{\stackrel{{\scriptstyle \rm k}}{\rm k \neq j}}\
\tilde{f}_{\rm jk} } }} + {1 \over 2}
\sum_{\stackrel{{\scriptstyle \rm j}}{\rm j \neq i}}\
  \tilde{g}_{\rm ji} \right] + \epsilon_{\rm i0} \quad ,
$$
where $\epsilon_{\rm i0}$ is independent of $\vec{\tilde{p}}_{\rm i}$.
This can be written as
$$
\epsilon_{\rm i} = \sqrt{ \left( m^*_{\rm i} \right)^2 +
\vec{\tilde{p}}_{\rm i}^2 } + \epsilon^{\prime}_{\rm i0} \quad ,
$$
where the effective mass $m^*_{\rm i}$ turns out to be
\begin{equation}
\label{eq:msr}
m^*_{\rm i} = 1
  - {1 \over 2}
\sum_{\stackrel{{\scriptstyle \rm j}}{\rm j \neq i}}\
  {\tilde{f}_{\rm ij} \over{
    1 + 2 {\displaystyle
  \sum_{\stackrel{{\scriptstyle \rm k}}{\rm k \neq j}}\
  \tilde{f}_{\rm jk} } }}
  - {1 \over 2}
  \sum_{\stackrel{{\scriptstyle \rm j}}{\rm j \neq i}}\
   \tilde{g}_{\rm ji} \quad .
\end{equation}
\begin{figure}[t]
\vspace*{-10mm}
\insertplot{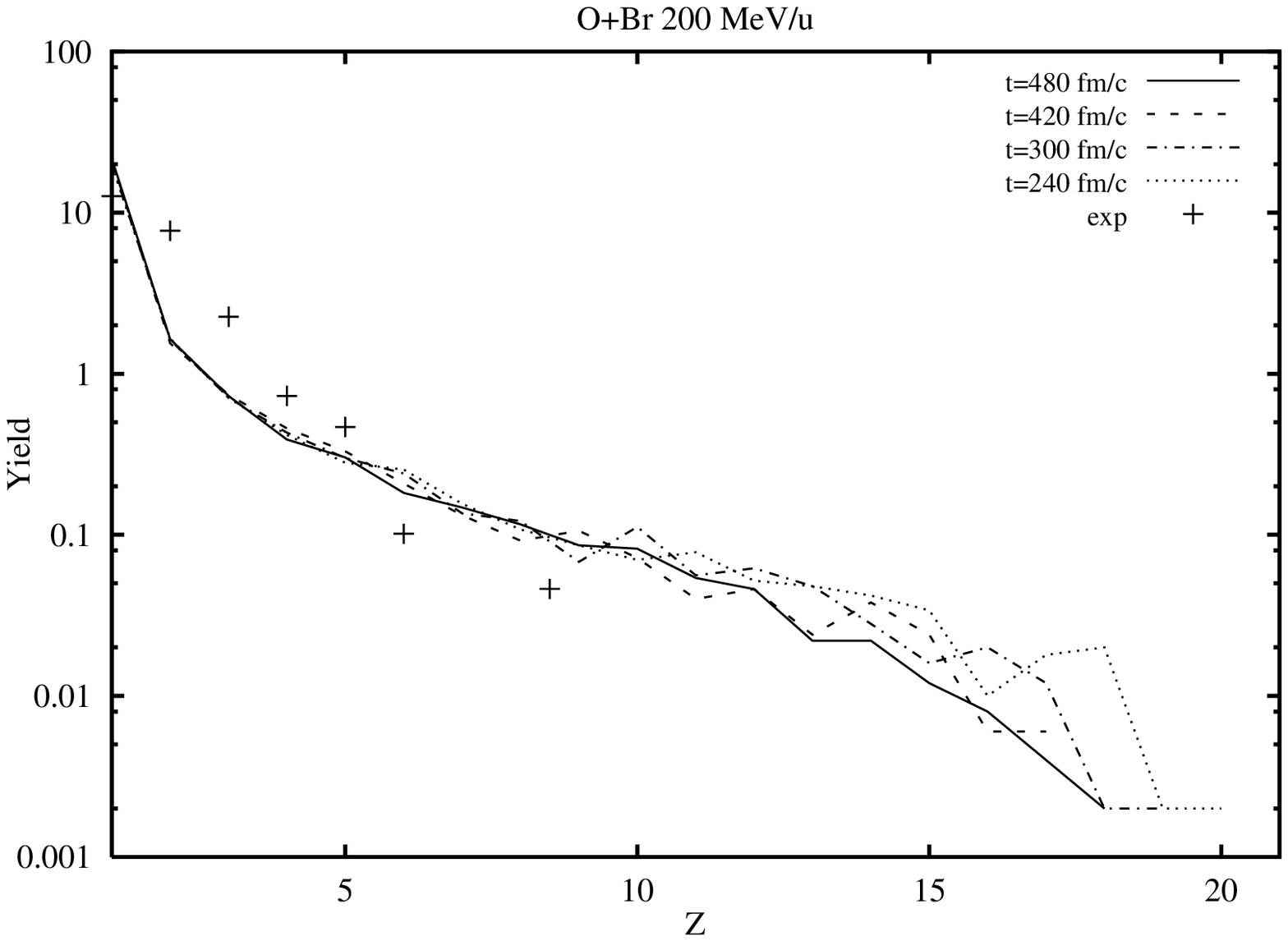}
\vspace*{-10mm}
\caption{The evolution of the fragment distribution in time in 200 MeV/u
O + Br central collisions}
\end{figure}

Case B: If we assume, that the momentum dependence of
$\tilde{f}_{\rm ij}$ and $\tilde{g}_{\rm ji}$ is the same as in nuclear matter,
that is
\begin{equation}
E = \sum_{\rm i}\
    \sqrt{ 1 + \vec{\tilde{p}}_{\rm i}^2 }
  - {1\over 2} \sum_{\stackrel{{\scriptstyle \rm i,j}}{\rm j \neq i}}\
    {f(r_{\rm ij}) \over{ \sqrt{ 1 + \vec{\tilde{p}}_{\rm i}^2 }
        \sqrt{ 1 + \vec{\tilde{p}}_{\rm j}^2 } }}
  { 1 \over { 1 + 2 {\displaystyle
    \sum_{\stackrel{{\scriptstyle \rm k}}{\rm k \neq j}}\
    f(r_{\rm jk}) } }}
  + {1\over 2} \sum_{\stackrel{{\scriptstyle \rm i,j}}{\rm j \neq i}}\
    g(r_{\rm ji}) \quad ,
\end{equation}
(see Eq.~(\ref{eq:nms})),then the effective mass for finite systems turns out
to be
\begin{equation}
\label{eq:msi}
m^*_{\rm i} = 1 -
  \sum_{\stackrel{{\scriptstyle \rm j}}{\rm j \neq i}}\
  {f(R_{\rm ij}) \over{
    1 + 2 {\displaystyle
  \sum_{\stackrel{{\scriptstyle \rm k}}{\rm k \neq j}}\
    f(R_{\rm jk}) } }} \quad .
\end{equation}
We found in our calculations that the definition~(\ref{eq:msr}) gave better
conservation of the total energy.

\subsection*{\large\sl 3.5. The treatment of a relativistic collision}
\begin{figure}
\vspace*{-10mm}
\insertplot{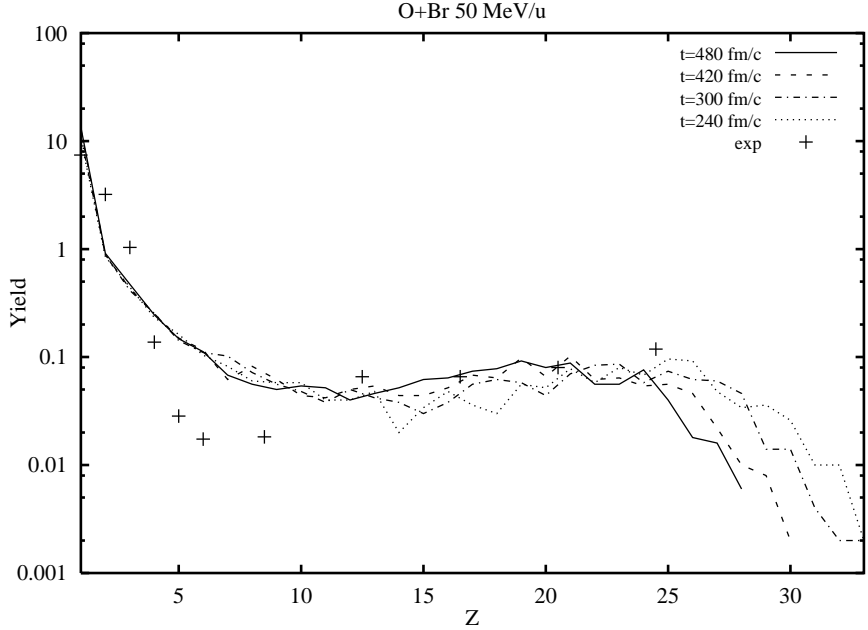}
\vspace*{-10mm}
\caption{The evolution of the fragment distribution in time in 50 MeV/u
O + Br central collision}
\end{figure}

The individual two-body collision is calculated in a relativistically covariant
way, as given in
\cite{kn:wolf}. The
particles are allowed to collide
if the impact parameter
\begin{eqnarray}
b &=& \sqrt{R_{12}^2 - h_{12}^2/v_{12}^2} \qquad \mbox{with} \\
R_{12}^2 &=& -(x_1 - x_2)^2 + \left( {p_1 (x_2 - x_1) \over{ m_1 c}} \right)^2
\quad , \nonumber \\
h_{12} &=& {p_1 (x_2 - x_1) \over{ m_1 c}} - {p_2 (x_2 - x_1) m_1 c \over{
p_1 p_2 }} \quad \mbox{and} \nonumber \\
v_{12}^2 &=& 1 - \left( {m_1 m_2 c^2 \over{ p_1 p_2 }} \right)^2 \nonumber
\end{eqnarray}
is smaller than the collision radius $\sqrt{\sigma/\pi}$. In
the model the collision takes place at equal time in the CM of the total
system, provided the closest approach of the world-lines occurs within the
following time step. After the collision we only allow final momenta which do
not let the particles approach each other further.

\subsection*{\large\sl 3.6. Pauli blocking}

Pauli blocking is treated as in Ref.~\cite{kn:judqmd}.
We calculate the averaged phase-space distribution function at the location
of each particle. The collision can occur if the new phase-space distribution
function is smaller than 1
for the colliding particles and for all other particles after the collision.
Since the distribution function at the phase space location of a particle
depends on the momenta of all the other particles, without this condition it
may occur that the phase
space density becomes greater than one even for particles which did not
take part in the collision.

\subsection*{\large\sl 3.7. Definition of fragments}

We consider the particles to belong to the same fragment if they stay together
in coordinate and momentum space and are bound. In our
calculation we follow the evolution of the system up to 360 fm/c for higher,
and up to 720 fm/c for lower energies. We see in the Figs.~4 and 5
that in the last 120 fm/c the so determined fragment distribution does
not change anymore.

In order to calculate the energy of a cluster we calculate the $f_{\rm ij}$
and $g_{\rm ij}$ in the total CM frame because they are Lorentz-scalars. Only
the momenta appearing in the momentum dependent terms have to be Lorentz
transformed to the center of momentum frame of the cluster.
The radii and the energies of the
fragments are very near to the experimental values, although some fragments can
shrink a little. However, we do not get energies lower than $-8.8$ MeV/u
for any case.

\begin{table}[t]
\begin{center}
\begin{tabular}{||c|c|c|c|c||} \hline
 & Scalar & Vector & $\vec{p}_{\rm i} \vec{p}_{\rm j}$ term & 3. order term \\
\hline \hline
$\varrho_0$ & -153 & 122 & -0.01 & 2.2 \\
1 MeV/u & -56 & 28 & 0.005 & 0.5 \\ \hline
3 $\varrho_0$ & -647 & 529 & 0.3 & 38 \\
1 MeV/u & -90 & 59 & 0.04 & 2.19 \\ \hline
$\varrho_0$ & -110 & 175 & -85 & 17 \\
2 GeV/u & -40 & 40 & -19 & 14 \\ \hline
3 $\varrho_0$ & -235 & 381 & -169 & 80 \\
2 GeV/u & -66 & 89.7 & -39 & 51.8 \\ \hline
\end{tabular}
\end{center}
\caption{The contribution of the different energy terms of
\protect{Eq.~(\ref{eq:en})} in MeV per particle
for gold on gold systems for very extreme cases. The upper numbers refers to
the Walecka, the lower ones to the ZM forces. The calculations
are made for 1 MeV/u and 2 GeV/u energies at normal and 3 times
normal nuclear matter density.}
\end{table}

\section*{\large\bf 4. Multifragmentation in the O\ +\ Br collision}

Among the experiments carried out to study nuclear multifragmentation, emulsion
measurements were the first to give an almost exclusive
identification of the atomic numbers of the fragments emitted in heavy ion
collisions.
The investigation of the same system at several bombarding energies is
particularly useful for testing models. Such data are available for the O + Br
system at bombarding energies 50, 75, 100, 150 and 200 MeV/u \cite{kn:jakob}.
Nowadays modern experimental techniques allow mass production of such
measurements \cite{kn:FOPI}. Few years ago we
carried out calculations for the emulsion experiment with a
non-relativistic force to test the treatment of the Pauli
blocking~\cite{kn:judqmd}. Now we have made the analogue calculation with
relativistic forces at different energies to check their applicability at lower
energies. These forces are meant to be used at higher energies, but they should
also have good non-relativistic properties.
\begin{figure}
\vspace*{-10mm}
\insertplot{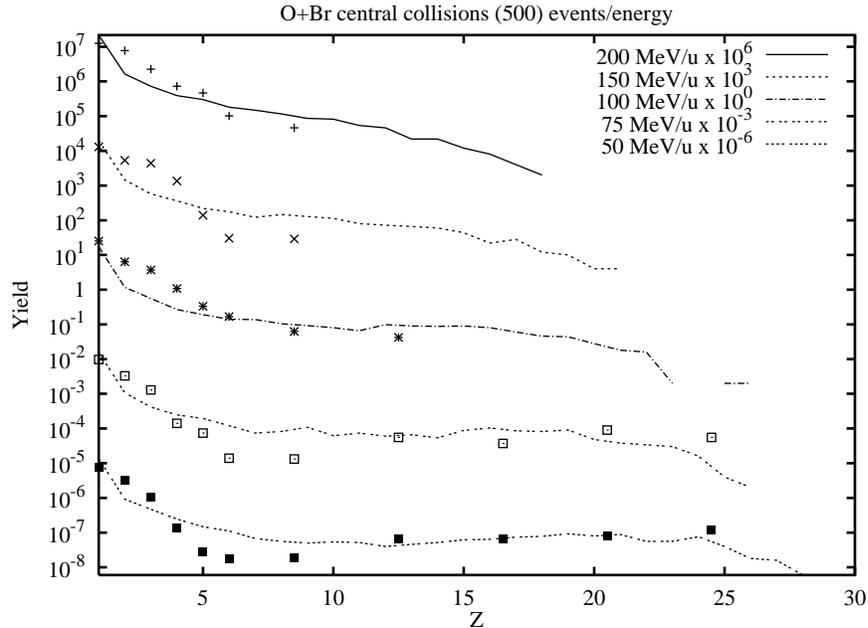}
\vspace*{-10mm}
\caption{Fragment distribution in O + Br central collisions at energies 50,
75, 100, 150 and 200 MeV/u}
\end{figure}

As a first step we checked the validity of the expansion in $f_{\rm ij}$ and
$g_{\rm ij}$. In Table~2 we give the values of different energy terms of
Eq.~(\ref{eq:en}) for a gold on gold system which was artificially compressed
to different densities and given different relative energies.
One can see, that for the ZM case the contribution of the
meson fields are much smaller than the nuclear rest mass $mc^2$. Even for the
Walecka case, except maybe for the high compression with $j^0=3 \varrho_0$, the
expansion is applicable. Thus
our approximation in the expansion (\ref{eq:mom}) of the momentum is justified.

In Fig.~6 the fragment distributions of the O + Br collisions are displayed for
different energies. The contribution from the
largest fragments are very small. Due to the computer time needed, the figures
present results averaged over only 500 events. Thus the statistical fluctuation
for events with small yields are still large.
In Fig.~7 we compare the results obtained
with the Walecka and ZM forces using the Cugnon parameterization \cite{kn:cug}
and our fitted cross sections for 50 MeV/u and 200 MeV/u bombarding energies.
The ZM force gives a somewhat better agreement with data than the forces
derived from the Walecka lagrangian. The effect is larger for higher energies.
However, using the Cugnon
cross section gives more or less the same results as using the one fitted
to the low energy free nucleon-nucleon scattering.

The fragment distributions shown in Fig.~6 do not differ significantly from
the earlier ones calculated with non-relativistic Skyrme forces, however
one should mention, that for lower energies the experimental cut does not
correspond to the central events. In Fig.~8 we display the distribution of
impact parameters for the 10\% highest multiplicities. For
the 200 MeV/u collision the multiplicity cut selects the central events (impact
parameter is less than 3 fm), but for 50 MeV it results in broad distribution.
The yields for 50 MeV/u and 200 MeV/u beam energies with the multiplicity cut
are given in Fig.~9.

\section*{\large\bf 5. Conclusion}
\begin{figure}
\vspace*{-10mm}
\insertplot{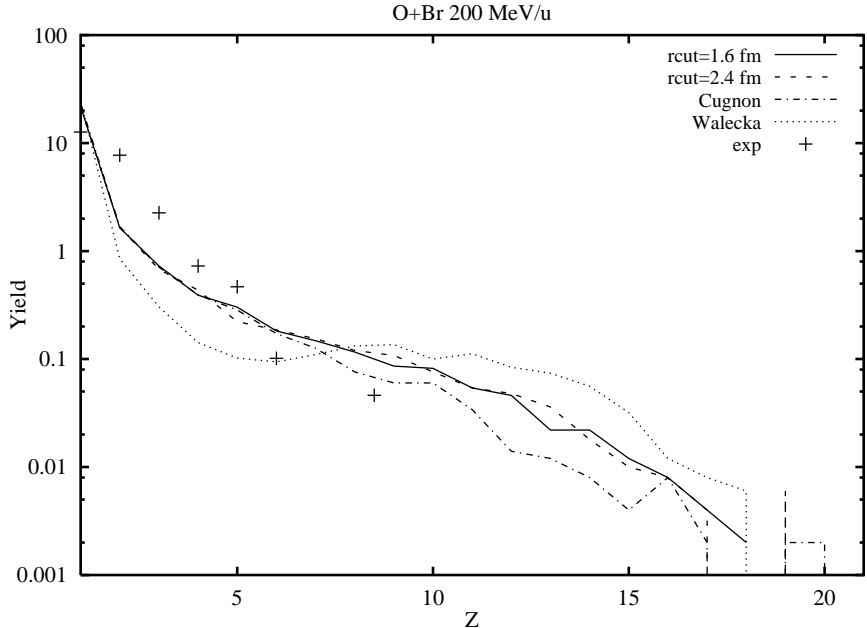}
\vspace*{-10mm}
\caption{Fragment distribution in O + Br central collisions at 200 MeV/u for
different cross sections and forces}
\end{figure}

{}From an effective field theory lagrangian with scalar and vector mesons
we deduced in the small acceleration approximation
coordinate and momentum dependent relativistic forces,
which are suitable for transport
equation calculations. Even without an explicit
density dependence, saturation of nuclear matter is obtained at proper
densities and energies
due to relativistic effects. The Zim\'anyi-Moszkowski forces, where a
derivative coupling is used, can produce saturation even without
relativistic effects.

The proposed forces will be used in the intermediate energy domain
(500 MeV/u -- 2 GeV/u), where the mean-field plays still an important role and
relativistic forces are necessary. However, they should work
at lower energies too. As a first check we studied their low energy
behavior in a QMD calculation. We investigated both, the more static features
of these forces in the ground-states of nuclei and their dynamical
properties in the fragmentation process in heavy ion collisions.
The fragment distributions in the O + Br (50 -- 200 A MeV) collision were
calculated. The result we obtained
is of similar quality as the one with the non-relativistic
Skyrme forces~\cite{kn:judqmd,kn:paula}.

That encourages us to use these forces for beam energies in the domain of
500 MeV/u -- 2 GeV/u, where non-relativistic Skyrme type forces cannot be used
anymore.
A further test of the model should be for example the comparison of
the calculated flow (which is especially sensitive to the momentum
dependence) for different energies with experimental results of the
FOPI collaboration~\cite{kn:FOPI}.
Such calculations are under progress.
\begin{figure}
\vspace*{-10mm}
\insertplot{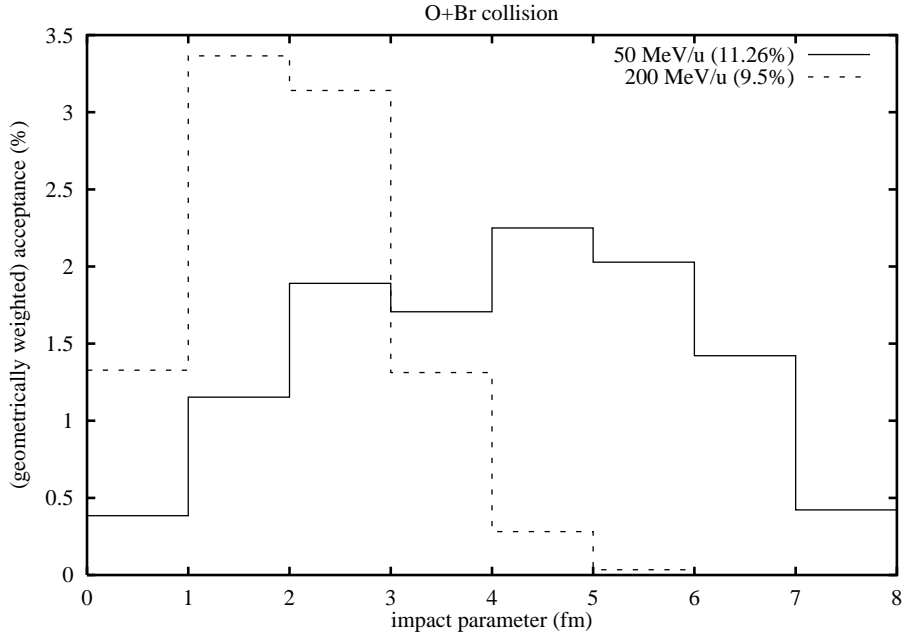}
\vspace*{-10mm}
\caption{The distribution of impact parameters to the
highest multiplicities. The solid line corresponds to 50 MeV/u collisions
and selects 11.3\% of the events, while the dashed line represents
9.5\% for 200 MeV/u}
\end{figure}

\begin{figure}
\vspace*{-10mm}
\insertplot{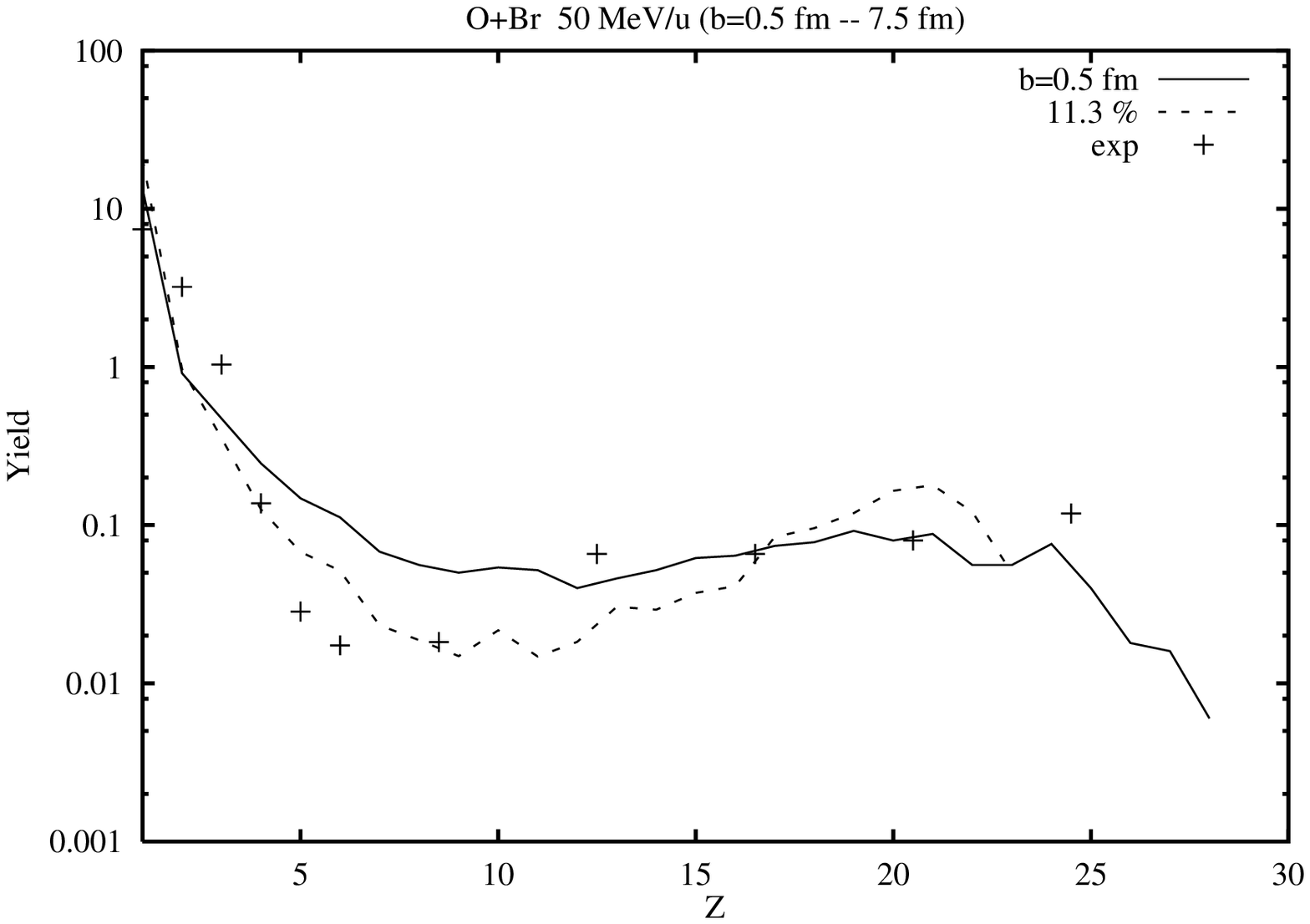}
\vspace*{-10mm}
\insertplot{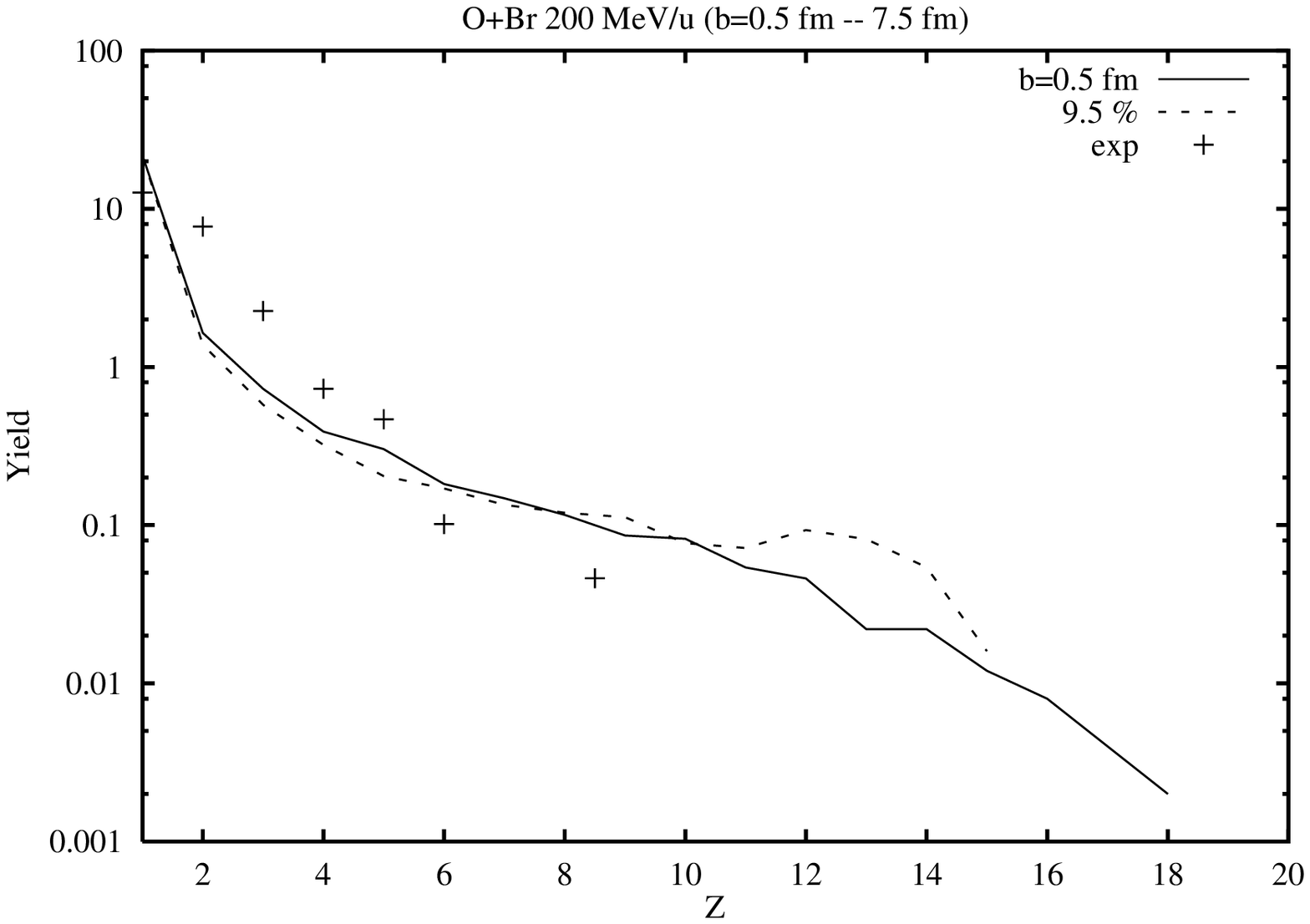}
\caption{The yields for 50 MeV/u (upper) and 200 MeV/u (lower) beam energies
with trigger to the events with the 10\% highest multiplicities. The solid line
displays the yield obtained for a central collision (b=0.5 fm), the dashed line
displays the average over the highest multiplicity events.
The experimental data are shown by crosses}
\end{figure}

\section*{\large\bf Acknowledgments}

One of the authors (J.N.) should like to express her thanks to Prof.
N\"orenberg
for his kind hospitality at the GSI, where part of this work was done. G.P.
thanks Gy. Wolf for fruitful discussions.
This work was supported in part by the Hungarian Research Foundation OTKA.

\section*{\large\bf Notes}

E-mail: H. Feldmeier: H.Feldmeier@gsi.de ; J. N\'e\-meth:
judit@hal9000.elte.hu ;\\
 G. Papp: G.Papp@gsi.de; WWW:http://www.gsi.de/\verb+~+papp .\\
$^\dagger$: Permanent address

%\newpage
%\begin{center}References\end{center}
%\vspace*{-15mm}

\end{document}